\documentclass[twocolumn]{aastex62}

\graphicspath{{./}{figures/}}

\received{}
\revised{}
\accepted{}

\submitjournal{ApJ}

\shorttitle{}
\shortauthors{}

\usepackage{epstopdf}
\usepackage{color}
\usepackage{graphicx}

\begin{document}

\title{\large The Chromospheric Response to the Sunquake generated by the X9.3 Flare of NOAA 12673}

\author{Sean Quinn}
\affil{Astrophysics Research Centre, School of Mathematics and Physics, Queen's University Belfast, BT7 1NN, Northern Ireland, UK \\
}

\author{Aaron Reid}
\affil{Astrophysics Research Centre, School of Mathematics and Physics, Queen's University Belfast, BT7 1NN, Northern Ireland, UK \\
}

\author{Mihalis Mathioudakis}
\affil{Astrophysics Research Centre, School of Mathematics and Physics, Queen's University Belfast, BT7 1NN, Northern Ireland, UK \\
}

\author{Christoper Nelson}
\affil{Astrophysics Research Centre, School of Mathematics and Physics, Queen's University Belfast, BT7 1NN, Northern Ireland, UK \\
}

\author{S. Krishna Prasad}
\affil{Astrophysics Research Centre, School of Mathematics and Physics, Queen's University Belfast, BT7 1NN, Northern Ireland, UK \\
}

\author{Sergei Zharkov}
\affil{E.A. Milne Centre for Astrophysics, School of Mathematics and Physical Sciences, Hull University, Kingston upon Hull, HU6 7RX, England, UK}

\begin{abstract}

Active region NOAA 12673 was extremely volatile in 2017 September, producing many solar flares, including the largest of solar cycle 24, an X9.3 flare of 06 September 2017. It has been reported that this flare produced a number of sunquakes along the flare ribbon \citep{Sharykin2018, Zhao2018}. We have used co-temporal and co-spatial Helioseismic and Magnetic Imager (HMI) line-of-sight (LOS) and Swedish 1-m Solar Telescope observations to show evidence of the chromospheric response to these sunquakes. Analysis of the Ca II 8542 \AA\space line profiles of the wavefronts revealed that the crests produced a strong blue asymmetry, whereas the troughs produced at most a very slight red asymmetry. 
We used the combined HMI, SST datasets to create time-distance diagrams and derive the apparent transverse velocity and acceleration of the response. These velocities ranged from 4.5 km s$^{-1}$ to 29.5 km s$^{-1}$ with a constant acceleration of 8.6 x 10$^{-3}$ km s$^{-2}$. We employed NICOLE inversions, in addition to the Center-of-Gravity (COG) method to derive LOS velocities ranging 2.4 km s$^{-1}$ to 3.2 km s$^{-1}$. Both techniques show that the crests are created by upflows. We believe that this is the first chromospheric signature of a flare induced sunquake. 
\end{abstract}


\section{Introduction} 
\label{sec:intro}

Solar flares involve the impulsive release of energy throughout the solar atmosphere. They are associated with active regions with large polarity inversion lines where the complexity of the magnetic field gives rise to reconnection in the corona. The anti-parallel movement of charged plasma can lead to the formation of a `current sheet' and the instability created causes opposing filaments to reconnect. The plasma above the reconnection site can be accelerated into interplanetary space while the plasma below moves down the filaments and towards the solar surface. It is this downward moving plasma that can disturb the lower atmosphere. The suggestion that solar flares can generate seismic response was first put forward by \cite{Wolff} and was more recently explored by \cite{KosovichvZharkova1995}. The first detection of flare induced SQs was reported in 1998 \citep{KosovichevZharkova1998}, using the Michelson Doppler Interferometer (MDI) \citep{Scherrer1995} onboard the Solar and Heliospheric Observatory (SOHO) \citep{Domingo1995}. The SQs appeared as expanding circular waves in photospheric Dopplegrams that were created during the impulsive phase of the flare event.

The widely accepted explanation for the generation of the SQ is associated with the `collisional thick-target' model (CTTM), where a SQ is produced by acoustic waves that travel into the solar interior where they are refracted by the denser plasma. In this model, a beam of high-energy particles is accelerated from the reconnection site towards the deeper layers of the solar atmosphere where it deposits large amounts of energy and momentum \citep{Kosovichev2007}. The collisions of the accelerated particles with the chromospheric plasma result into heating and the production of hard and soft X-Ray (HXR and SXR) emission. The heated plasma expands, causing a compression known as a `chromospheric condensation', which travels deeper into the solar atmosphere and collides with the denser photosphere. The collision creates a high pressure compression in the photosphere, causing a downward propagating shock front \citep{Kosovichev2006}. The shock imparts energy through the convection zone, where it is reflected due to changes in density and temperature, and appear as expanding ripples on the solar surface.

Alternative sources of energy deposition into the photosphere have also been proposed. These include photospheric heating due to continuum radiation \citep{lindsey_and_donnea2008}, or deeply penetrating proton beams \citep{ZharkovaZharkov2007}. These models can explain seismic events where HXR or white light (WL) emission is detected and provide an explanation for SQ generation. One of the issues with these interpretations is that the shock front which propagates in the solar interior can experience significant damping, depleting the energy of the seismic wave \citep{Russell2016}. Furthermore, the source of the SQ is sometimes detected away from the site of the HXR emission \citep{Zharkov2011}. This suggests that the SQ is not uniquely formed at the site of the accelerated particle collisions. The SQ ripples are often observed during the impulsive phase of the flare, before the maximum HXR and SXR emission \citep{Zharkov2011}. 

While SQs are sometimes observable using Dopplergrams alone, time-distance diagrams can also be employed for the analysis of the observed ridges \citep{KosovichevZharkova1998, ZharkovaZharkov2007, Sharykin2018, Zhao2018}. Once the time-distance diagram is created, a theoretical regression trend can normally be fitted to the data, with a match to any SQ ridge that is present in the diagram. The regression trend uses the ray-path approximation \citep{couvidat} with a more detailed description provided by \cite{Kosovichev2011}.

Further analysis can be conducted by implementing regression techniques such as acoustic holography \citep{DoneaLindsey2005, Donea2011}. Acoustic holography can be used to reveal the source of the SQ by creating a partial reconstruction of the acoustic field using Dopplergrams of the flaring region. These images are averaged over their cadence and convolved with a Green's function, in terms of height in the atmosphere, as well as horizontal distance from an approximate source and time \citep{Doneaetal1999}. The Green's function allows the observed ripple in the Dopplergrams to be traced back to a point source \citep{Matthews2011, Matthews2015}. The convolution of these two functions creates a regression map. The regression power map displays sources of acoustic emission as bright kernels, and sources of acoustic absorption as dark kernels \citep{Doneaetal1999}. The map can be filtered in frequency to gain information on the formation height of the sources and sinks.

\begin{figure*}
\plotone{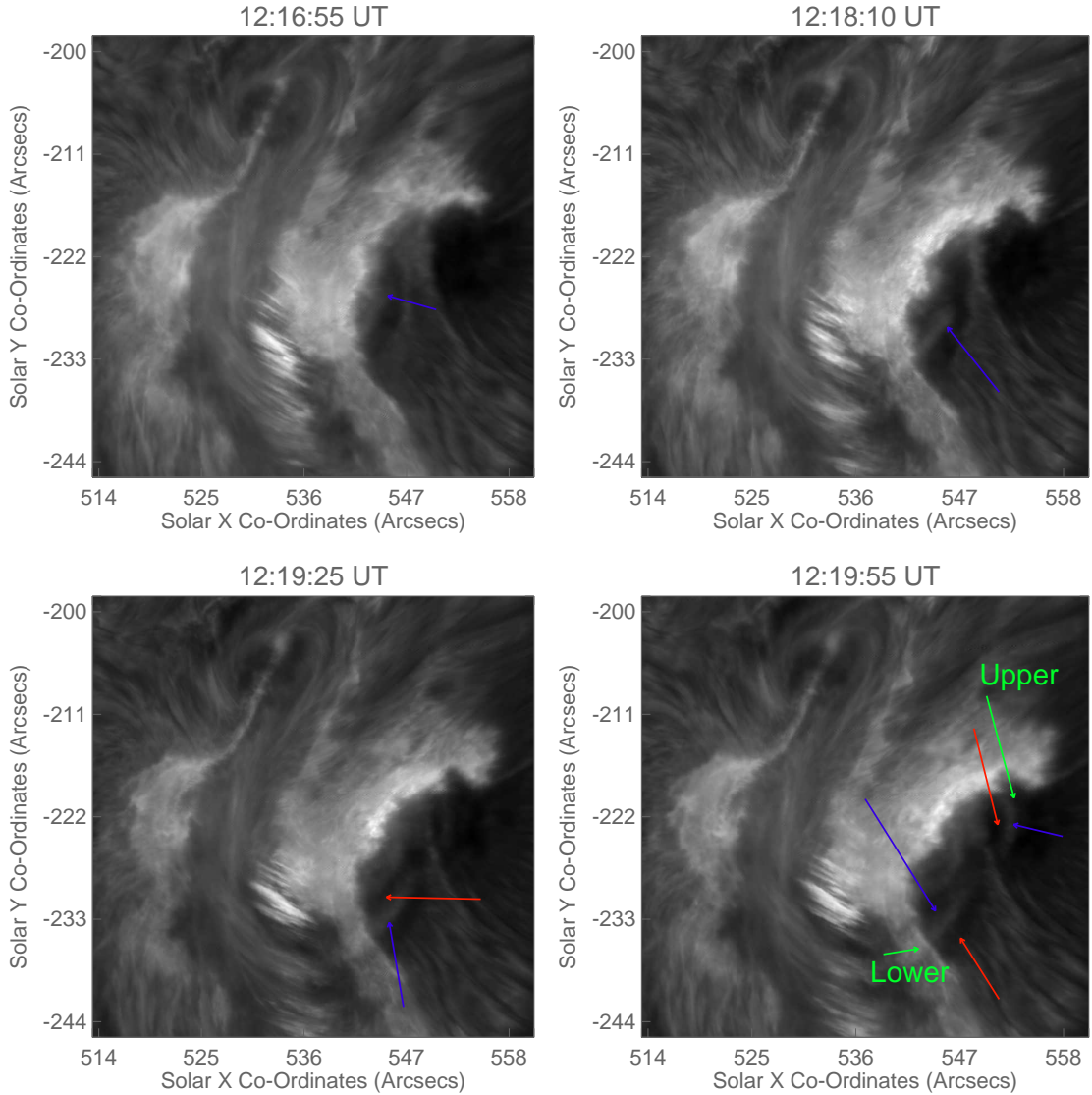}
\caption{The 47\arcsec$\times$47\arcsec\ SST FOV analysed here plotted at four time-steps (indicated by each panel title) at 8541.8 \AA. The blue arrows point to `crests' (brightenings) and the red arrows point to `troughs' (darkenings) of the chromospheric response to the SQ. The two green arrows in the bottom right panel point to the 'upper' and 'lower' responses.}
\label{Fig1}
\end{figure*}

Despite the increased number of SQ detections provided by the Solar Dynamics Observatory (SDO) \citep{Pesnell} their relationship with the flare energy released remains unclear. For example, a number of X-class flares have been reported without SQs, where low M-class flares have \citep{Kosovichev-Helioseimology}. This is not to say that there is no seismic response present as it may lie below the solar background noise. Analysis of the atmospheric response to SQs, such as time delays and intensity variations, can provide insights on the temperature gradients and intensity variations in the solar interior. It has been predicted that a global seismic response can also be excited by a solar flare. Such seismic responses would have a very low amplitude, below the amplitude of random solar oscillations, hence detection has been proven difficult \citep{Kosovichev-Helioseimology}.

In this article, we present evidence of a chromospheric response to a sunquake following the X9.3 solar flare of 2017 September 6, the largest flare of Solar Cycle 24. The flare originated from Active Region (AR) NOAA 12673 and the GOES flux peaked at approximately 11:55:00 UT. Photospheric disturbances associated with this flare have been reported in the literature (\citealt{Sharykin2018, Zhao2018}). We apply time-distance analysis to derive the apparent transverse velocity and acceleration of the SQ. The corresponding LOS velocity is derived with the COG method and inversion techniques.

\section{Observations} 
\label{sec:Observations}

\begin{figure*}
\plotone{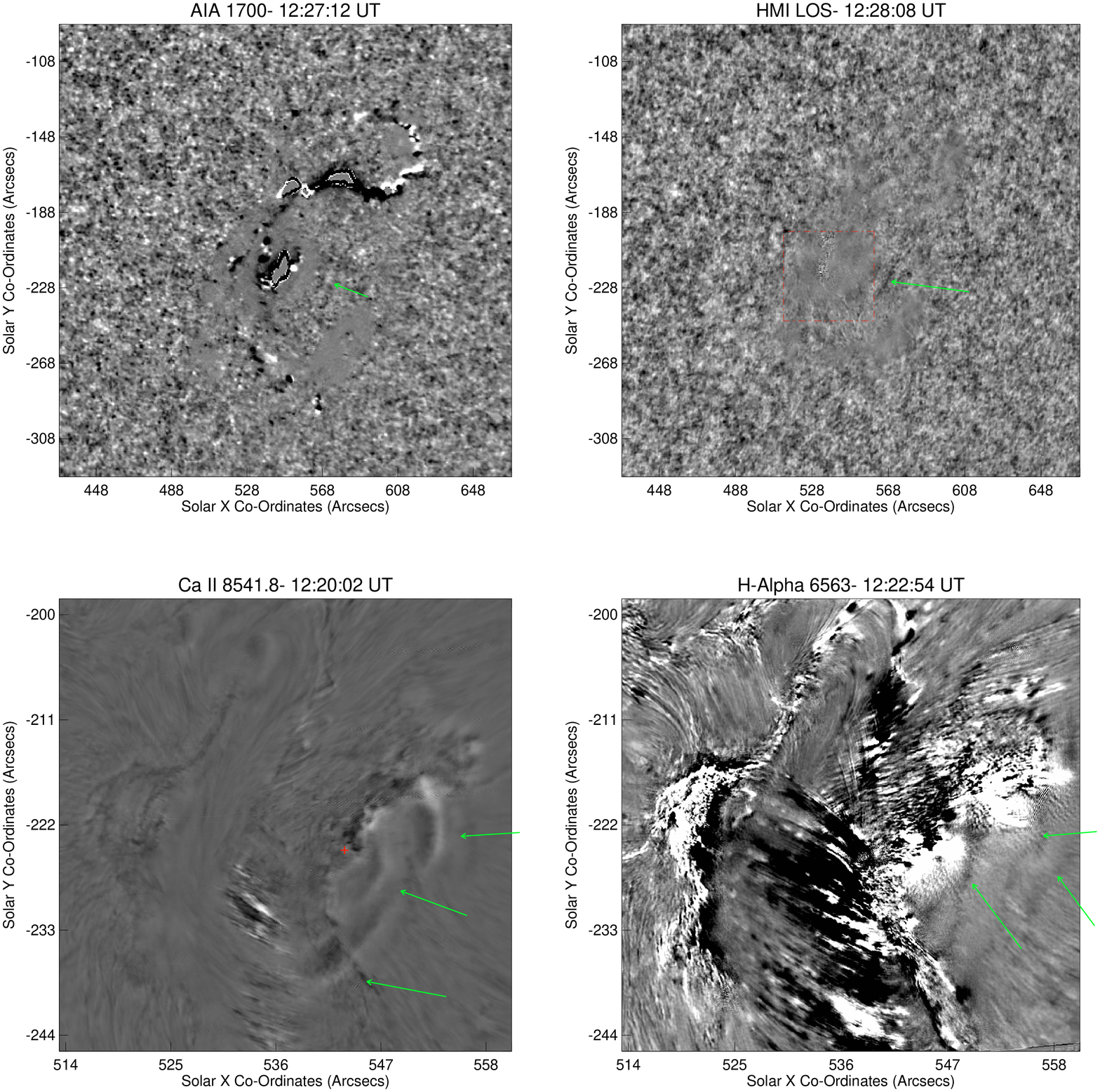}
\caption{Difference images that show the seismic response of the atmosphere to the X9.3 flare of 06 September 2017. Top: SDO AIA 1700 \AA\ and the HMI LOS running differences. The dotted red box over-laid on the HMI LOS image indicates the CRISP FOV plotted in the bottom panels. Bottom: Co-temporal SST running difference images from Ca II 8541.8 \AA\ and H$\alpha$. Green arrows indicate the presence of wavefronts. The red cross in the bottom left image is the estimated epicentre of the Ca II 8541.8 \AA\ response.\\
(An animation of this figure is available.)\\}
\label{Fig2}
\end{figure*}

The ground-based data analysed here were collected between 11:57:17 UT and 12:52:05 UT on 2017 September 6 by the CRisp Imaging SpectroPolarimeter (CRISP; \citealt{Scharmer2008}) attached to the Swedish Solar Telescope (SST; \citealt{Scharmer2003}). During this time, CRISP was pointed at coordinates of X=537\arcsec, Y=-222\arcsec\ ($\mu$=0.79) and ran a sequence sampling the \ion{Ca}{2} 8542 \AA\ and H$\alpha$ lines, with the \ion{Ca}{2} 8542 \AA\ data being observed in full-Stokes polarimetry mode. The \ion{Ca}{2} 8542 \AA\ scan included 11 line positions at $\pm0.7$ \AA, $\pm0.5$ \AA, $\pm0.3$ \AA, $\pm0.2$ \AA, $\pm0.1$ \AA, as well as the line core. The H$\alpha$ scan included 13 line positions, at $\pm1.5$ \AA,\space $\pm1.0$\AA,\space $\pm0.8$ \AA,\space $\pm0.6$ \AA, $\pm0.3$ \AA, $\pm0.15$ \AA, and the line core. Wide-band (WB) images were also obtained co-temporal with each CRISP narrow-band image for alignment purposes. 

These SST data were processed using the Multi-Object Muti-Frame Blind Deconvolution (MOMFBD; \citealt{Lofdahl, vanNoort}) method. This included the sub-division of each image into 88x88 pixel$^{2}$ sub-images, each of which was individually restored to maintain the presumed invariance of the image formation modules. A pre-filter FOV and wavelength dependent correction was applied to each restored image. More information on the SST reduction pipeline is provided in \cite{SST-Pipline}. The analysis detailed in this article was conducted on a reduced CRISP FOV of 47.0\arcsec$\times$47.0\arcsec\ in order to avoid fringing effects at the edge of the CCDs. These data have a pixel scale of $0.059$\arcsec\ and cadence of 15 s. 

In addition to CRISP, we also analyse data sampled by the Solar Dynamics Observatory's Helioseismic and Magnetic Imager (SDO/HMI; \citealt{Scherrer12}) and Atmospheric Imaging Assembly (SDO/AIA; \citealt{Pesnell}). HMI obtains full disk line-of-sight (LOS) velocity and magnetic field measurements, as well as photospheric continuum images, every 45 s with a pixel scale of $0.5$\arcsec. A detailed explanation on the HMI LOS velocity maps is provided in \cite{HMI, HMI_LOS}. The HMI LOS velocity data best displayed the presence of a SQ, and was chosen to be the photospheric representation for this analysis. It should be noted, though, that the HMI continuum and magnitude data cubes also showed some seismic responses.

Only the 1600 \AA\ and 1700 \AA\ filters from the SDO/AIA instrument were analysed here. These filters clearly displayed the SQ and had a pixel scale of 0.6\arcsec\ and a cadence of 24 s. An extended 240\arcsec$\times$240\arcsec\ FOV was studied for all SDO data to investigate the propagation of the SQ analysed out of the limited FOV of the CRISP instrument. The SDO data were co-aligned with the SST using routines developed by R. J. Rutten\footnote{http://www.staff.science.uu.nl/$\sim$rutte101/rridl/sdolib/}.

\section{Results}

\subsection{Properties of the SQs}

During the evolution of the flare, the \ion{Ca}{2} 8542 \AA\ data revealed multiple wavefronts emanating from the flare ribbon. In Fig.~\ref{Fig1}, we plot four temporal snapshots of the flare and associated SQ at a spectral position of 8541.8 \AA\ (0.2 \AA\ away from the \ion{Ca}{2} 8542 \AA\ line core). This line position was chosen as it showed the clearest signature of the SQs. The two green arrows in the bottom right image indicate the seismic responses, appearing as wavefronts, that emanated from 2 sources in the flare ribbon. The `upper' source appeared to excite waves propagating westward, while the `lower' source emitted waves moving in a south-westerly direction. We refer to these sources as `upper' and `lower' sources due to their locations in relation to the flare ribbon. The blue and red arrows highlight crests and troughs of the apparent waves, respectively. It should be noted that the wavefronts can be detected at other position across the \ion{Ca}{2} 8542 \AA\ line profile, however, the contrasts between the waves and the background were not as great at these spectral locations as at 8541.8 \AA.

To better accentuate these seismic responses, we created running difference maps for both the SDO and CRISP \ion{Ca}{2} 8542 \AA\ datasets. These running difference maps were created using the formuala {\it diff}(n)={\it frame}(n)-{\it frame}(n-1). The chromospheric \ion{Ca}{2} 8542 \AA\ seismic response to the flare was first detected in these running difference images at approximately 12:11:33 UT. Although wavefronts could also be detected in a running difference of the H$\alpha$ line core, the flare ribbon was significantly brighter in these data meaning achieving accurate inferences about the SQ properties proved more difficult. Therefore, the majority of this work was undertaken on the \ion{Ca}{2} 8542 \AA\ data. Running difference maps from four diagnostics are shown in Fig.~\ref{Fig2}, where the green arrows over-laid on each panel indicate the wavefronts. The red box on the HMI LOS image (top right) depicts the co-temporal SST CRISP FOV plotted in the lower panels. The first seismic response appears in the HMI Dopplegram at approximately 12:12:02 UT from the lower source. As this response was observed to be the longest, it will be the main focus of our subsequent analysis.

Analysis of the co-aligned CRISP and HMI LOS datasets revealed that the \ion{Ca}{2} 8542 \AA\ seismic wave signatures propagated in the same direction as those detected in HMI (Figure 3). The blue arrow indicates the wave detected in CRISP data at 12:18:37 UT, and the red arrow indicates the HMI LOS wave at 12:22:57 UT. While there are multiple SQs detected in HMI, the wavefront that aligns both spatially and temporally with the chromospheric response first appeared at 12:12:02 UT. The correspondence between the signatures in CRISP and HMI data will be studied in more detail in the following sub-section. As the CRISP instrument has a smaller FOV and higher spatial resolution than the HMI, it is not unexpected that the chromospheric signature of the SQ is detected prior to the photospheric signature. The stark intensity contrast in the H$\alpha$ running difference discussed previously can be seen in the bottom right of Figure 2. Any intensity increase created as a wavefront moves across the image would be lost in the apparent `background noise' created by the contrasting intensities.

\begin{figure*}
\plotone{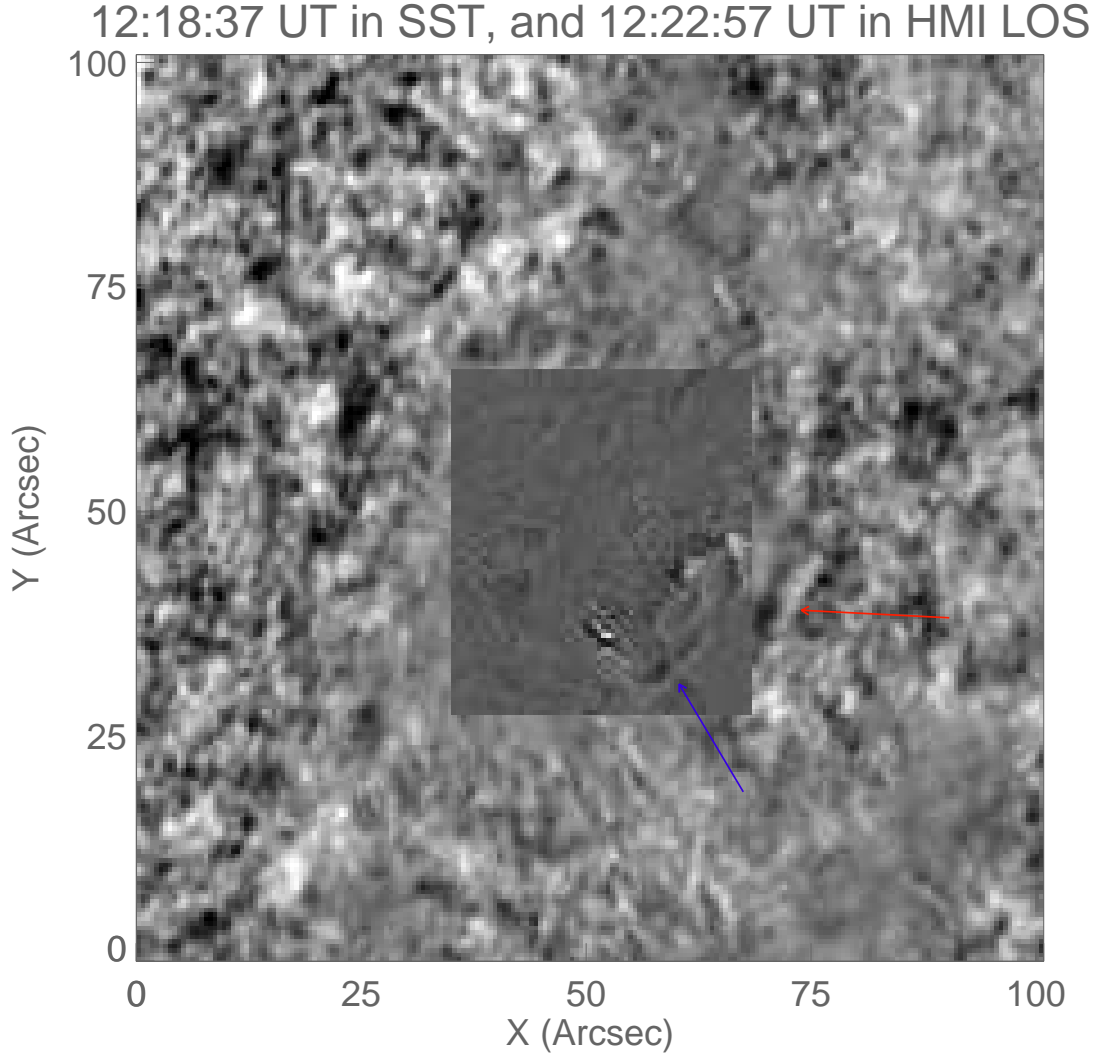}
\caption{A Ca II 8542 \AA\ running difference (over-laid panel) aligned to a HMI LOS running difference image (background panel) showing the progression of the wavefront across the two datasets. The CRISP image was obtained at 12:18:37 UT and the HMI at 12:22:57 UT. The blue and red arrows indicate the seismic response in the chromosphere and photosphere, respectively.}
\label{Fig3}	
\end{figure*}

\subsection{Time-distance analyses}

\begin{figure*}
\plotone{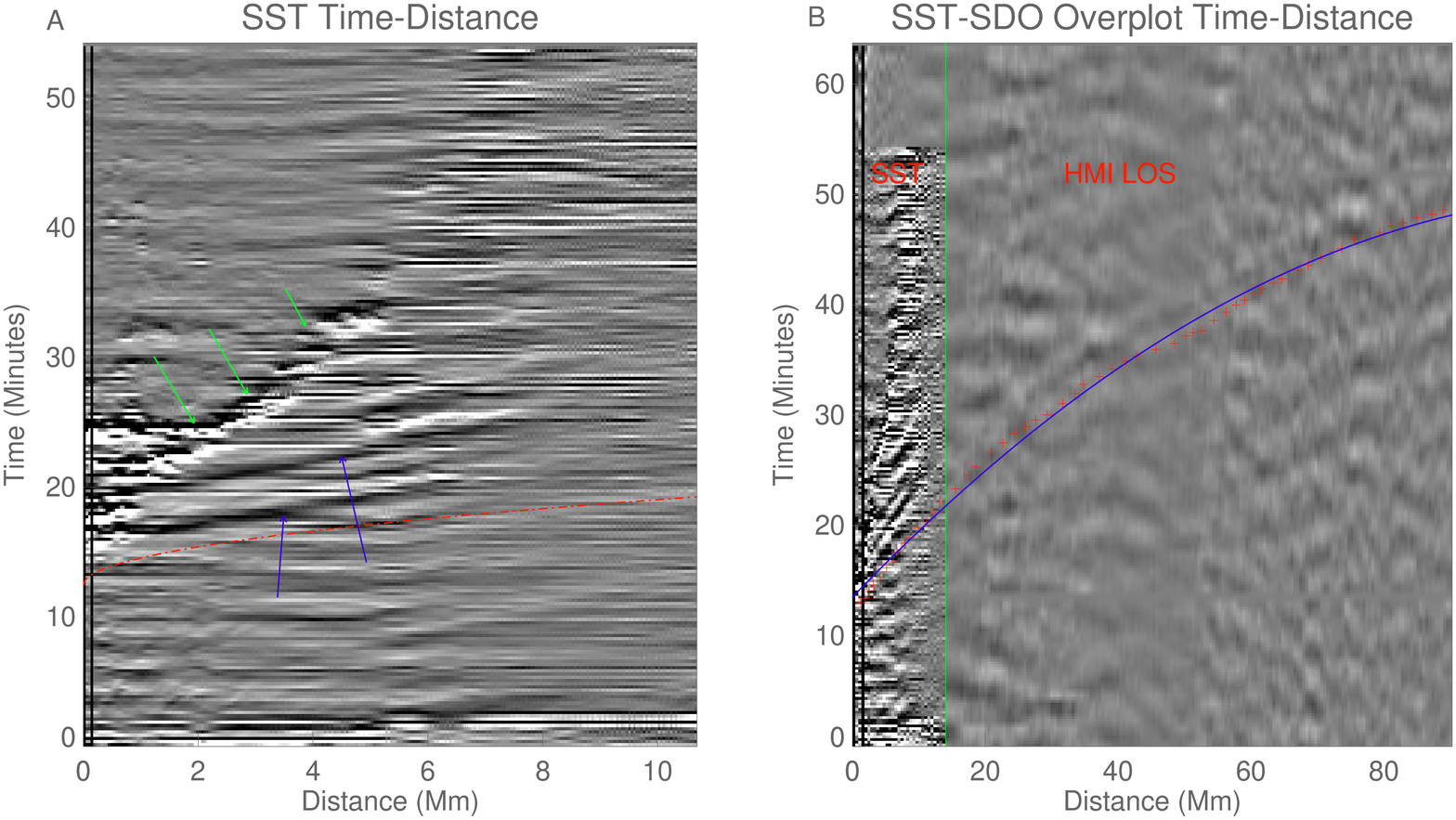}
\caption{Left (A): Time-distance diagram generated using CRISP running difference imaging at $8541.8$ \AA. The red dotted line shows the ridge formed by the first detected wavefront. This line has been shifted vertically by a few pixels. The blue arrows indicate the ridges created by the wavefronts. The ridges are very slightly curved, similar to the regression trend of $x^{0.5}$ that is expected from seismic responses. The upward propagating structure to the left of the image, indicated by the green arrows, is the noise created as the flare ribbon moves into the area used to create the time-distance diagram. Right (B): Time-distance diagram created using the combined CRISP-HMI running difference data cube. The multiple ridges that can be seen in the CRISP part of the diagram correlate to the multiple wavefronts of the HMI data. One of these ridges extends out of the CRISP FOV and matches well to the ridge in the HMI LOS part of the diagram. The red crosses on this image are points selected by hand, and used to perform the $\chi^2$ fit analysis, and the blue plot is the fit that most accurately matches these points. Both the red points and the blue plot have been shifted vertically to ease the visibility of the ridge.}
\label{Fig4}
\end{figure*}

In order to study the progress of the waves from the lower source, an approximate point-of-origin was identified for these apparent waves and time-distance diagrams were created by selecting a pixel range and angle, over which the wavefront propagated. By tracing back the wavefronts, the origin of the seismic response was approximated as X=543.16\arcsec, Y=-225.87\arcsec, displayed as the red cross in the bottom left image of figure 2. Acoustic Holography \citep{Doneaetal1999} techniques were undertaken to better identify the sources of the seismic responses but these proved unsuccessful. This was most likely due to the difference in height at which the holography was undertaken. Normally, acoustic holography is undertaken in the photosphere, where ours was in the chromosphere. This difference on height most likely proved the problem, causing the technique to be unsuccessful. The waves were observed to propagate approximately $8.58$ Mm from this epicentre and between angles 125$^\circ$ and 145$^\circ$, where 0$^\circ$ is solar north. The left panel of Fig.~\ref{Fig4} plots the time-distance diagram constructed from the CRISP \ion{Ca}{2} 8541.8 \AA\ data. Pixels where the wavefront was present appeared more intense, hence they created ridges of higher intensity. Pixels in the trough of the wave were lower in intensity creating the striking light-dark streaks. The same procedure was undertaken for the upper response but this was more noisey and had more interference with the flare ribbon.

Multiple ridges can be seen in the CRISP time-distance diagram, indicated by the blue arrows in Fig.~\ref{Fig4}, consistent with the presence of multiple wavefronts in Fig.~\ref{Fig1}. We followed a similar procedure to that conducted by \citet{Sharykin2018} and fitted the initial ridge on the CRISP time-distance diagram in Fig.~\ref{Fig4} with a regression trend of $x^{0.5}$, that SQs commonly possess (\citealt{KosovichevZharkova1998}). We selected this ridge as it was not disturbed by the presence of the dynamic flare ribbons indicated with the green arrows on the left hand panel of Fig.~\ref{Fig4}. This theoretical trend matches well with the observed propagation of the wave and is depicted as the red dash-dotted line in the left-hand panel of Fig.~\ref{Fig4} (moved vertically downwards by a few pixels in order that it did not obscure the wavefront).

The observed Ca II 8542 \AA\ wavefronts, which move in an apparent circular arc pattern from two different locations, are consistent with being the chromospheric components of the photospheric SQs analysed by \cite{Sharykin2018}. In order to study the propagation of the waves from the CRISP FOV to the HMI FOV, a time-distance diagram that combined the CRISP and HMI datasets was constructed. This is plotted in image B of Fig.~\ref{Fig4}. The HMI data were extrapolated such that their cadence was co-temporal with the CRISP data and the spatial resolution of the CRISP dataset was degraded to that of HMI. The same epicentre and angle over which the wavefront propagated through as in image A of Fig.~\ref{Fig4} were used, however, a longer slit was created in order to study the wave propagation into the surrounding atmosphere. The left side of the CRISP-HMI time-distance diagram (to the left of the green vertical line) shows the `overplotted' CRISP data, with a number of ridges, which correspond to the multiple wavefronts observed in running difference. The times and positions where the wavefront can be identified in the combined CRISP-HMI dataset, correlate well with the position of the ridges in the CRISP time-distance diagram indicating that the seismic wave observed in the CRISP data is a chromospheric response to the photospheric SQ reported previously by \citet{Sharykin2018}.

Using the larger FOV provided by the combined CRISP and HMI dataset, we are able to further analyse the temporal behaviour of these seismic responses. The red dots in Figure 4(b) mark the apparent locations of the wavefront through time which are used to model the apparent propagation of the response. A $\chi^2$ analysis was then conducted to determine trend would best model these data-points: linear; a trend of $x^{0.5}$ (as theory would predict); or a second order polynomial. Our analysis revealed that a function of the shape of the theoretical regression trend, $x^{0.5}$, provided the most accurate fit to both the SST and HMI regions of the time-distance diagram. This trend is plotted by with the blue line on image B of Fig.~\ref{Fig4}. As in image A, these points have been shifted slightly downwards to allow the ridge to become more visible. 

\begin{figure*}
\plotone{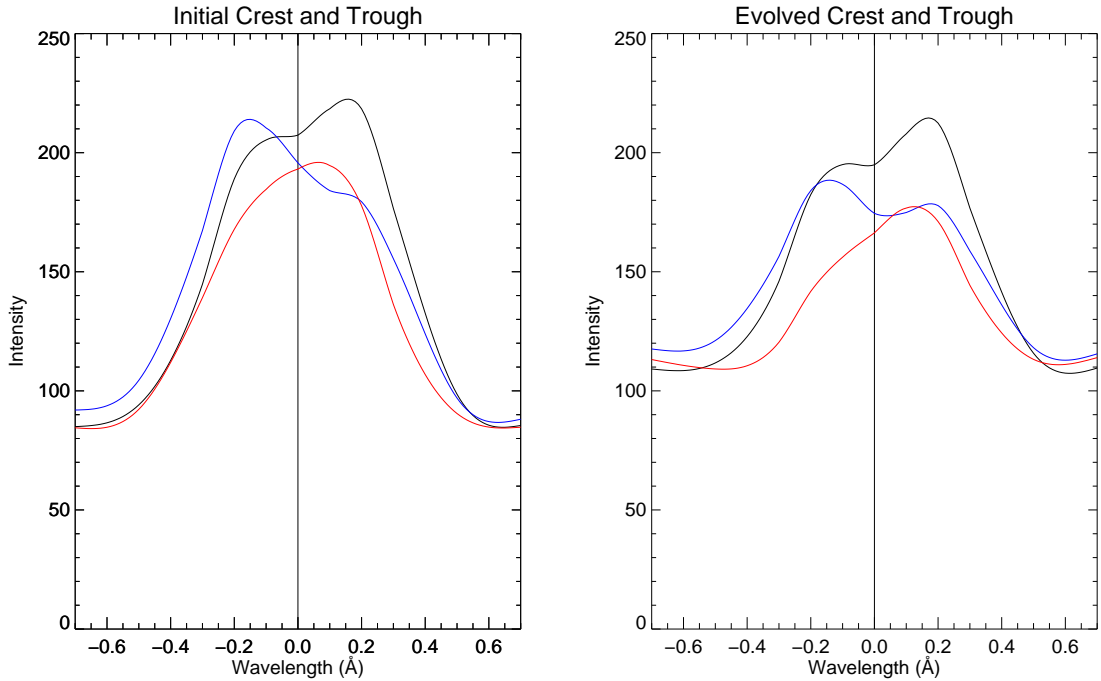}
\caption{An example of the line profiles created for each wavefront. A total of four wavefronts were sampled. The 'initial' line profiles are shown on the left and the `evolved' line profiles are shown on the right. Quiet line profiles are plotted in black whereas crests and troughs are shown in blue and red, respectively. The initial line profiles were taken early in this wavefront propagation, and the evolved profile was taken after the wavefront has had time to move across the FOV.}
\label{Fig5}
\end{figure*}

Tracking the apparent position of the wave through time allows us to estimate its velocity and acceleration. The velocity of the wavefront ranged from 4.5 km s$^{-1}$ at its initial time to 29.5 km s$^{-1}$ as it propagated out of the analysed FOV. The wavefront gradually accelerated at a rate of 8.6x10$^{-3}$ km s$^{-2}$ and travelled for approximately 117 Mm from the estimated epicentre in 46.5 minutes. This entire propagation was not used in the creating of figure 4.B, as the ridge becomes nearly indistinguishable from background noise at these distances, while remaining visible by eye in the HMI LOS running difference.

Before the calculation of these wavefront velocities, one may be reminded of another flare initiated chromospheric response, namely, Moreton waves \citep{Moreton, Chen2011}. These waves propagate across very large distances (5x10$^5$ km) at extremely high velocities (500-2000 km s$^{-1}$) \citep{Chen2011}. However, these velocities and propagation distances confirm the wavefronts are a representation of a chromospheric component of a SQ.

 A number of such wavefronts are apparent in the HMI LOS running difference data, however, only one appears to propagate outside of the SST FOV. We have also investigated the SQ in the SDO/AIA 1600\AA\ and 1700\AA\ channels. Our selection for a region of interest in these channels was guided by HMI and the cadence was adjusted to fit with the HMI running difference. Signatures of the SQ were detected in both AIA channels. An AIA 1700\AA\ running difference map is also shown in the top left panel of Figure 2. The SQ is less pronounced in AIA 1600\AA. The AIA wavefronts are detected further away from the flare ribbon and appear later than in HMI. We investigated these AIA channels in an attempt to follow the upwards propagating response to the SQ, between the HMI LOS and the 8542 \AA\ observations. The AIA 1600 \AA\ and 1700 \AA\ observations were the only channels that detected the response. These channels are sensitive to the upper photosphere and photosphere, respectively \citep{AIA}. We also investigated the AIA 304 \AA\ observations, which are sensitive to the high-chromosphere/transition region \citep{AIA}, to determine if the response was able to propagate to these heights, however no such response was detected.

\subsection{LOS velocity of the SQs}

\begin{figure*}
\plotone{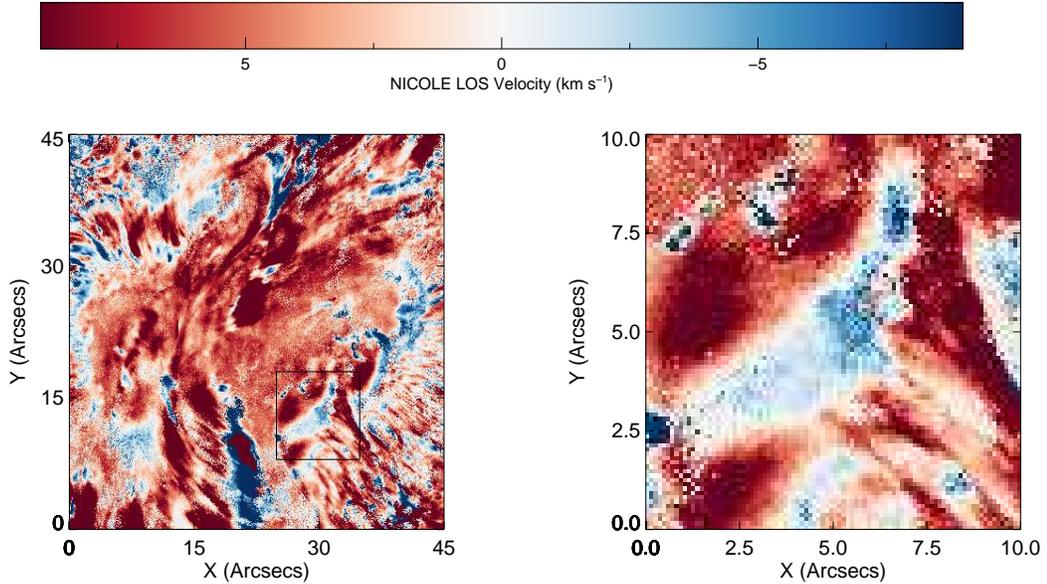}
\caption{LOS velocity maps created using the NICOLE Inversion algorithm on the Ca II 8542\AA\ data for the full FOV (left panel) and a zoomed in region around the wavefront (right panel). The black box indicates the FOV plotted in the right-hand panel. The NICOLE velocity map has been created for a log $\tau_{5000} = -3$.} 
\label{Fig6}
\end{figure*}

We have also analysed the spectral profiles co-spatial to these apparent wavefronts using Ca II 8542\AA\ data. The line profiles were investigated to determine the level of any line asymmetries. Each wavefront was subdivided into 20x20 pixel$^2$ areas for computational ease. A line profile was generated from each of the boxes, and the profiles were combined to create an average for each crest and trough. The same procedure was applied as the crests and troughs moved across the FOV. An average `quiet' profile was created from the same spatial location before it was crossed by the wavefront. The quiet profiles were averaged over 20 consecutive frames that did not show any intensity variations or velocity shifts to minimise the influence of the nearby sunspot. 

In order to quantify the line asymmetries, we generated intensity ratios between the blue and red peaks. The ratios were created by integrating the line flux within $\pm$0.1\AA\space of the maximum for the blue (I$_B$) and the red (I$_R$) peaks respectively. These peaks were found by fitting the line profiles with two single Gaussians, one for the red peak and one for the blue. The same procedure was applied to the quiet profiles. In Fig.~\ref{Fig5}, asymmetries in the \ion{Ca}{2} 8542\AA\ line profiles created from crests and troughs are clearly present. The initial line profile was constructed from crests approximately 4 Mm from the epicentre and the evolved profiles were sampled after the crest had propagated to 5 Mm from the epicentre. These distances were selected as they were the best positions to find a wavefront that was significantly far from the ribbon, minimising any intensity contamination. Any further from this chosen distance, the wavefronts tended to interfere with each other or the ribbon, making the creation of a single evolved wavefront line profile impossible. Poor seeing was also a constraint in regards to which frames were of a good enough quality to create a reliable line profile. Both the initial and evolved profiles displayed the same trend, with the crest showing a blue asymmetry and the trough with a slight red asymmetry. 

Normalising the crests and troughs to their quiet profiles allowed any underlying asymmetries to be negated. Four sets of crests and troughs were investigated, and the initial and evolved normalised asymmetry ratios are displayed in Table 1. Following normalisation against their quiet profiles, the trough line profiles had an I$_{B}$/I$_{R}$ ratio of $\approx$1. The trends shown by the line asymmetries provide evidence that the seismic response creates a velocity perturbation.

\begin{table*}
	\centering
	\begin{tabular}{|c||c|c|c|c|}
\hline
& Initial Crest & Initial Trough & Evolved Crest & Evolved Trough\\
\hline
\hline
Wave 1& 1.22 & 0.99 & 1.25 & 0.95\\
\hline
Wave 2& 1.22 & 1.00 & 1.22 & 1.04\\
\hline
Wave 3& 1.24 & 0.94 & 1.23 & 1.06\\
\hline
Wave 4& 1.18 & 1.05 & 1.18 & 0.99\\
\hline
\end{tabular}
\caption{Table of normalised asymmetry ratios of I$_{B}$/I$_{R}$ for each line profile.}
\end{table*}

We used the NICOLE inversion algorithm to create a LOS velocity  map of the propagating seismic response. NICOLE solves the multi-level NLTE radiation transfer problem for each pixel by following the preconditioning method set out in \cite{navarro}. It perturbs a number of parameters such as temperature, magnetic field, electron density, micro-turbulence and LOS velocity of an initial atmosphere to find the best match with the observations \citep{SocasNavarro2000}. The pre-initial atmosphere that was used in this case is the Harvard-Smithsonian Reference Atmosphere (HSRA) model \citep{hsra}. The CRISP observations were interpolated to ensure even spacing between each spectral point with weighting assigned only to the observed spectral positions. A 2x2 binning was applied across the entire image. Each line profile was normalised with a number of reference profiles, taken from a `quiet' frame, across a multitude of regions with no major flare disturbances. The results of this step were used as the initial guess atmosphere for the subsequent inversions.

The data were then inverted to obtain an initial estimate of the atmosphere with very little vertical stratification. The inversion used 4 nodes in temperature, 1 node in LOS velocity, 1 node in micro-turbulence, and 1 node in LOS magnetic flux density. The output was smoothed spatially in three dimensions, and this was used as input in the next iteration. This cycle involved 7 nodes in temperature, 3 in LOS velocity, 3 in LOS magnetic flux density, 1 in transverse magnetic flux density, and 1 node in micro-turbulence. A third cycle was attempted, with increased nodes in both magnetic flux density and velocity, but this did not show significant improvement from the second cycle. Figure 6 shows the velocity maps created using NICOLE inversions for the entire FOV (left hand panel) and for a zoomed region around the apparent wavefront. The wavefronts in the lower chromosphere at an optical depth of log$\tau_{5000\AA}=-3$ display clear upflows, which is to be expected given the CTTM of SQ generation. 

The NICOLE outputs were compared with the centre of gravity (COG) method \citep{Uitenbroek} that can also be used to calculate Doppler velocities. Given this is the first detection of the chromosphere responding to a SQ, we wanted to estimate the corresponding LOS velocities with different, independent methods. The average LOS velocity of the Ca II 8542\AA\ wave indicated in Fig.~\ref{Fig6} found using the COG method is 2.39 km s$^{-1}$ while the corresponding velocity value determined from NICOLE is 3.20 km s$^{-1}$. This difference can be accounted for by the simplicity of the COG method, however, it is below the velocity accuracy we can obtain with our spectral sampling. Therefore, we assume that the upflow velocity is approximately $3.2$ km s$^{-1}$, given that NICOLE is more intricate than the COG method, we give it a higher weighting. As a comparison, we also analysed the HMI LOS velocity co-spatial to the wavefronts. The photospheric LOS velocity was approximately $2$ km s$^{-1}$, meaning it is consistent with the chromospheric velocities.

\section{Conclusions}

We investigated the seismic responses in the solar photosphere and chromosphere generated by the X9.3 GOES class solar flare of 2017 September 06. The co-spatial and co-temporal analysis of imaging and imaging spectroscopy obtained from HMI, AIA in the photosphere and CRISP in the chromosphere allowed us to conclude that photospheric SQs do have signatures in the chromosphere. Numerous wavefronts are apparent in time-distance diagrams constructed from difference imaged CRISP data, however, only one wavefront is apparent in HMI LOS velocity maps. This wavefront matches well to one wavefront from CRISP. The apparent propagation of this wavefront's plane-of-sky velocity was calculated to be increasing from 4.5 km s$^{-1}$ initially to 29.5 km s$^{-1}$ after 46.5 minutes.

We have used the COG method and NICOLE inversions of the Ca II 8542\AA\space line to construct LOS velocity maps. The LOS velocities of the upflowing material increase as they move through the atmosphere. The average upflow velocity detected in the photosphere is 2 km s$^{-1}$, with this increasing to 3.2 km s$^{-1}$ in the lower chromosphere. 

 Interestingly, upper-chromospheric observations using the AIA 304 \AA\ channel show no evidence of the SQ. This may mean the seismic response was unable to propagate to this height. However, this may also be due to the blooming effects due the saturation of the flare ribbon which which are most pronounced between 11:54:30 UT and 12:47:07 UT. The magnetic canopy may also be inhibiting the propagation of the wavefront higher into the atmosphere. 

To the best of our knowledge this is the first detection of a flare generated SQ in the chromosphere. However, we have shown that the signal of such responses is very weak, and doesn't move far from the flare ribbon in the chromosphere. The lack of previous detections of such responses in the chromosphere is not due to the lack of a response being present, but due to low spectral and temporal resolutions in previous observational setups.

\begin{acknowledgements}
The Swedish 1m Solar Telescope is operated on the island of La Palma by the Institute for Solar Physics of Stockholm University in the Spanish Observatorio del Roque de los Muchachos of the Instituto de Astrofsica de Canarias. The Institute for Solar Physics is supported by a grant for research infrastructures of national importance from the Swedish Research Council (registration number 2017-00625). SQ acknowledges support from the Northern Ireland Department for the Economy for the award of a PhD studentship. AR, MM, CJN and SKP acknowledge support from STFC under grant No. ST/P000304/1.
\end{acknowledgements}


\begin{thebibliography}{}
\expandafter\ifx\csname natexlab\endcsname\relax\def\natexlab#1{#1}\fi
\providecommand{\url}[1]{\href{#1}{#1}}
\providecommand{\dodoi}[1]{doi:~\href{http://doi.org/#1}{\nolinkurl{#1}}}
\providecommand{\doeprint}[1]{\href{http://ascl.net/#1}{\nolinkurl{http://ascl.net/#1}}}
\providecommand{\doarXiv}[1]{\href{https://arxiv.org/abs/#1}{\nolinkurl{https://arxiv.org/abs/#1}}}

\bibitem[{Chen {et~al.}(2011)Chen, Ding, Chen, \& Harra}]{Chen2011}
Chen, F., Ding, M., Chen, P., \& Harra, L. 2011, The Astrophysics Journal, 470,
  116

\bibitem[{Couvidat {et~al.}(2004)Couvidat, Birch, Kosovichev, \&
  Zhao}]{couvidat}
Couvidat, S., Birch, A.~C., Kosovichev, A.~G., \& Zhao, J. 2004, The
  Astrophysical Journal, 607, 554

\bibitem[{Couvidat {et~al.}(2011)Couvidat, Schou, Shine, Bush, Miles, Scherrer,
  \& Rairden}]{HMI_LOS}
Couvidat, S., Schou, J., Shine, R., {et~al.} 2011, Solar Physics, 275, 285

\bibitem[{de~la Cruz~Rodr{\'\i}guez {et~al.}(2014)de~la Cruz~Rodr{\'\i}guez,
  L{\"o}fdahl, S{\"u}tterlin, Hillberg, \& Rouppe van~der Voort}]{SST-Pipline}
de~la Cruz~Rodr{\'\i}guez, J., L{\"o}fdahl, M., S{\"u}tterlin, P., Hillberg,
  T., \& Rouppe van~der Voort, L. 2014, Astronomy and Astrophysics, 573, A40

\bibitem[{Domingo {et~al.}(1995)Domingo, Fleck, \& Poland}]{Domingo1995}
Domingo, V., Fleck, B., \& Poland, A.~I. 1995, Solar Physics, 162, 1

\bibitem[{Donea(2011)}]{Donea2011}
Donea, A. 2011, Space Science Review, 158, 451

\bibitem[{Donea {et~al.}(1999)Donea, Braun, \& Lindsey}]{Doneaetal1999}
Donea, A.~C., Braun, D.~C., \& Lindsey, C. 1999, The Astrophysical Journal,
  513, L143

\bibitem[{Donea \& Lindsey(2005)}]{DoneaLindsey2005}
Donea, A.~C., \& Lindsey, C. 2005, The Astrophysics Journal, 630, L168

\bibitem[{Gingerich {et~al.}(1971)Gingerich, Noyes, Kalkofen, \& Cuny}]{hsra}
Gingerich, O., Noyes, R.~W., Kalkofen, W., \& Cuny, Y. 1971, Solar Physics, 18,
  347

\bibitem[{Kosovichev(2006)}]{Kosovichev2006}
Kosovichev, A.~G. 2006, Solar Physics, 238, 1

\bibitem[{Kosovichev(2007)}]{Kosovichev2007}
---. 2007, The Astrophysical Journal, 670, L65

\bibitem[{Kosovichev(2011)}]{Kosovichev2011}
---. 2011, in 'Pulsation of the {S}un and stars', Lecture Notes in Physics, 832

\bibitem[{Kosovichev(2014)}]{Kosovichev-Helioseimology}
---. 2014, in Extraterrestrial Seismology, 306

\bibitem[{Kosovichev \& Zharkova(1995)}]{KosovichvZharkova1995}
Kosovichev, A.~G., \& Zharkova, V.~V. 1995, in 'Helioseimology', Vol.~2,
  341--344

\bibitem[{Kosovichev \& Zharkova(1998)}]{KosovichevZharkova1998}
---. 1998, Nature, 393, 317

\bibitem[{Leman {et~al.}(2012)Leman, Title, Akin, Boerner, Chou,
Drake, Duncan, Edwards, Friedlaender, Heyman, Hulburt, Katz, Kushner, Leavay, Lindgren, Mathur, McFeaters, Mitchell, Rehse,  Schrijver, Springer, Stern, Tarbell, Wuelser, Wolfson, Yanari,  Bookbinder, Cheimets, Caldwell, Deluca, Gates, Golub,  Park,  Podgorski, Bush, Scherrer, Gummin, Smith, Auker, Jerram, Pool, Soufli, Windt, Beardsley, Clapp, Lang \& Waltham}]{AIA}
Leman, J., R., Title, A. , M., Akin, D., R., {et~al.} 2012, Solar Physics, 275, 17

\bibitem[{Lindsey \& Donea(2008)}]{lindsey_and_donnea2008}
Lindsey, C., \& Donea, A.~C. 2008, Solar Physics, 251, 627

\bibitem[{L{\"o}fdahl(2002)}]{Lofdahl}
L{\"o}fdahl, M.~G. 2002, in International Symposium on Optical Science and
  Technology, Vol. 4792, Image Reconstruction for Incomplete Data II, SPIE

\bibitem[{Long {et~al.}(2016)Long, Bloomfield, Chen, Downs, Gallagher, Kwon,
 Vanninathan, Veronig, Vourlidas, Vr{\v s}nak, Warmuth, \& {\v
 Z}ic}]{Long2016}
Long, D., Bloomfield, D., Chen, P., {et~al.} 2016, Solar Physics, 292, 1

\bibitem[{Matthews {et~al.}(2015)Matthews, Harra, Zharkov, \&
  Green}]{Matthews2015}
Matthews, S.~A., Harra, L.~K., Zharkov, S., \& Green, L.~M. 2015, The
  Astrophysical Journal, 812, 35

\bibitem[{Matthews {et~al.}(2011)Matthews, Zharkov, \& Zharkova}]{Matthews2011}
Matthews, S.~A., Zharkov, S., \& Zharkova, V.~V. 2011, The Astrophysical
  Journal, 739

\bibitem[{Moreton \& Ramsey(1960)}]{Moreton}
Moreton, G., E., \& Ramsey, H., E. 1960, Publications of the Astronomical
  Society of the Pacific, 72, 357

\bibitem[{Pesnell {et~al.}(2012)Pesnell, Thompson, \& Chamberlin}]{Pesnell}
Pesnell, W.~D., Thompson, B.~J., \& Chamberlin, P.~C. 2012, Solar Physics, 275,
  3

\bibitem[{Russell {et~al.}(2016)Russell, Mooney, Leake, \& Huson}]{Russell2016}
Russell, A. J.~B., Mooney, M.~K., Leake, J.~E., \& Huson, H.~S. 2016, The
  Astrophysical Journal, 831

\bibitem[{Scharmer {et~al.}(2003)Scharmer, Bjelksjo, Korhonenc, Lindbergd, \&
  Pettersonb}]{Scharmer2003}
Scharmer, G.~B., Bjelksjo, K., Korhonenc, T., Lindbergd, B., \& Pettersonb, B.
  2003, The 1-meter {S}wedish {S}olar {T}elescope, Vol. 162, 129--188

\bibitem[{Scharmer {et~al.}(2008)Scharmer, Narayan, Hillberg, de~la
  Cruz~Rodriguez, L{\"o}fdahl, Kiselman, S{\"u}tterlin, van Noort, \&
  Lagg}]{Scharmer2008}
Scharmer, G.~B., Narayan, G., Hillberg, T., {et~al.} 2008, The Astrophysical
  Journal Letters, 689, L69

\bibitem[{Scherrer {et~al.}(1995)Scherrer, Bogart, Bush, Hoeksema, Kosovichev,
  Schou, Rosenberg, Springer, Tarbell, Title, Wolfson, Zayer, \&
  Team}]{Scherrer1995}
Scherrer, P.~H., Bogart, R.~S., Bush, R.~I., {et~al.} 1995, Solar Physics, 162,
  129

\bibitem[{Scherrer {et~al.}(2012)Scherrer, Schou, Bush, Kosovichev, Bogart,
  Hoeksema, Liu, Duvall, Zhao, Title, Schrijver, Tarbell, \&
  Tomczyk}]{Scherrer12}
Scherrer, P.~H., Schou, J., Bush, R.~I., {et~al.} 2012, Solar Physics, 275, 207

\bibitem[{Schou {et~al.}(2011)Schou, Scherrer, Bush, Wachter, Couvidat,
  Rabello-Soares, Bogart, Hoeksema, Liu, Duvall, Akin, Allard, Miles, Rairden,
  Shine, Tarbell, Title, Wolfson, Elmore, Norton, \& Tomczyk}]{HMI}
Schou, J., Scherrer, P., Bush, R., {et~al.} 2011, Solar Physics, 275, 229

\bibitem[{Sharykin \& Kosovichev(2018)}]{Sharykin2018}
Sharykin, I.~N., \& Kosovichev, A.~G. 2018, Submitted to Astrophysics Journals,
  24 pages

\bibitem[{Socas-Navarro \& Bueno(1997)}]{navarro}
Socas-Navarro, H., \& Bueno, J.~T. 1997, The Astrophysical Journal, 490, 383

\bibitem[{Socas-Navarro {et~al.}(2000)Socas-Navarro, Trujillo~Bueno, \&
  Ruiz~Cobo}]{SocasNavarro2000}
Socas-Navarro, H., Trujillo~Bueno, J., \& Ruiz~Cobo, B. 2000, The Astrophysical
  Journal, 530, 977

\bibitem[{Uitenbroek(2003)}]{Uitenbroek}
Uitenbroek, H. 2003, The Astrophysical Journal, 592, 1225

\bibitem[{Van~Noort {et~al.}(2005)Van~Noort, Van Der~Voort, \&
  L{\"o}fdahl}]{vanNoort}
Van~Noort, M., Van Der~Voort, L.~R., \& L{\"o}fdahl, M.~G. 2005, Solar Physics,
  228, 191

\bibitem[{Wolff(1972)}]{Wolff}
Wolff, C.~L. 1972, The Astrophysics Journal, 176, 833

\bibitem[{Zhao \& Chen(2018)}]{Zhao2018}
Zhao, J., \& Chen, R. 2018, The Astrophysical Journal Letters, 860

\bibitem[{Zharkov {et~al.}(2011)Zharkov, Green, Matthews, \&
  Zharkova}]{Zharkov2011}
Zharkov, S.~I., Green, L.~M., Matthews, S.~A., \& Zharkova, V.~V. 2011, The
  Astrophysics Journal Letters, 741.2, L35 (6pp)

\bibitem[{Zharkova \& Zharkov(2007)}]{ZharkovaZharkov2007}
Zharkova, V.~V., \& Zharkov, S.~I. 2007, The Astrophysics Journal, 664, 573

\end{thebibliography}
\end{document}